\documentclass[letterpaper]{article}
\usepackage{aaai}
\usepackage{times}
\usepackage{helvet}
\usepackage{courier}
\usepackage{graphicx}
\usepackage{subfigure}
\usepackage{amsfonts,amsmath,amssymb}
\usepackage[utf8]{inputenc}

\begin{document}

\title{Privacy and the City: User Identification and Location Semantics in
Location-Based Social Networks}
\author{Luca Rossi, Matthew J. Williams, Christoph Stich, Mirco Musolesi \\ School of Computer Science \\ University of Birmingham \\ United Kingdom \\ \{l.rossi,m.j.williams,cxs489,m.musolesi\}@cs.bham.ac.uk}
\maketitle

\begin{abstract}
\begin{quote}
With the advent of GPS enabled smartphones, an increasing number of users is actively sharing their location through a variety of applications and services. Along with the continuing growth of Location-Based Social Networks (LBSNs), security experts have increasingly warned the public of the dangers of exposing sensitive information such as personal location data. Most importantly, in addition to the geographical coordinates of the user's location, LBSNs allow easy access to an additional set of characteristics of that location, such as the venue type or popularity. 

In this paper, we investigate the role of location semantics in the identification of LBSN users. We simulate a scenario in which the attacker's goal is to reveal the identity of a set of LBSN users by observing their check-in activity. We then propose to answer the following question: what are the types of venues that a malicious user has to monitor to maximize the probability of success? Conversely, when should a user decide whether to make his/her check-in to a location public or not? We perform our study on more than 1 million check-ins distributed over 17 urban regions of the United States. Our analysis shows that different types of venues display different discriminative power in terms of user identity, with most of the venues in the ``Residence" category providing the highest re-identification success across the urban regions. Interestingly, we also find that users with a high entropy of their check-ins distribution are not necessarily the hardest to identify, suggesting that it is the collective behaviour of the users' population that determines the complexity of the identification task, rather than the individual behaviour. 
\end{quote}
\end{abstract}
\medskip
\section{Introduction}
In the past years, the widespread use of Internet-connected smartphones capable of tracking our movements has had a significant impact on how we live our lives. The ubiquitous and constant connectivity has changed our habits, shaping the way we interact with other people, share information, perform tasks, and move around the city. Many services running on phones are built around the availability of location information. In particular, Location-Based Social Networks (LBSNs) bring together the geographic and the social dimensions, raising our awareness of the space surrounding us and allowing us to share recommendations on the venues we like or dislike with others. Foursquare\footnote{https://foursquare.com}/Swarm\footnote{https://www.swarmapp.com} is perhaps the most popular LBSN with more than 55 million users registered as of January 2015\footnote{https://foursquare.com/about/}. In addition to Foursquare, most Online Social Networks integrate location sharing features -- examples include Facebook\footnote{https://www.facebook.com/about/location} and Google+\footnote{https://support.google.com/plus/answer/2998354}.


The fundamental action that can be performed in a LBSN is a \emph{check-in}. Through a check-in, a user can register his/her position at a certain venue for friends and other users to see. Users can also leave feedback to describe their experience at a venue, which benefits other individuals who may be seeking recommendations while exploring their surroundings. 

The unrestrained sharing of personal location information raises serious concerns about the privacy of LBSN users. An increasing number of researchers have highlighted the dangers of exposing sensitive information such as location data~\cite{beresford2003location,bohn2005social,krumm2009survey,friedland2010cybercasing,rossi2014s}. For example, it has been shown that it is possible to identify individuals with a very high accuracy from a sample of their location data~\cite{de2013unique}. These privacy concerns are sometimes at odds with the design of LBSNs, which often encourage location sharing through gamification and incentives~\cite{ruiz2011location}. Therefore, privacy in LBSNs has been the focus of several studies~\cite{bettini2005protecting,gruteser2003anonymous,kalnis2007preventing,sweeney2002k,chow2011trajectory,ma2013privacy,pontes2012beware,pontes2012we,rossi2014s}.

The ability to link a LBSN service such as Foursquare with other social networks raises further privacy issues. A Facebook user who is not using Foursquare may inadvertently have their location revealed by another friend, since it is possible for Foursquare users to tag people they are connected to in Facebook~\cite{rossi2014s}.

In this paper, we investigate the interdependence between location semantics and privacy. More specifically, our goal is to determine if there exists a relation between the characteristics of a venue and the ability of an attacker to discriminate between the identities of different visitors of that venue. For example, check-ins at a user's home location are typically highly discriminative, as he/she is likely to be the sole user frequently visiting that location. On the other hand, information on the check-ins to a busy transportation hub or popular restaurant is less discriminative, since there are likely to be many different users with similar check-in patterns visiting popular locations. Intuitively, we would expect that there is heterogeneity in the discriminative power of venues in a LBSN, and this is one of the key questions we aim to answer in this paper. Furthermore, we are also interested in the extent to which the characteristics of a venue (e.g., type, popularity, location, etc.) are associated with its discriminative power. Such information is valuable to a malicious user seeking to reveal the identities of LBSN users from their check-in patterns, as it may inform the attacker on what types of venue he/she should monitor to maximize the probability of successfully identifying his/her victims. Finally, we also consider the related question of when should a user withhold his/her check-in to a particular venue so that they minimize the risk of being identified.

In order to answer the above questions, we simulate a scenario in which a malicious user infers other users' identities by monitoring their check-in behaviour. 
Our aim is to measure how the probability of re-identification success changes as we vary the attributes of the venues being monitored. More specifically, given a specific venue characteristic, e.g., venues in the ``Food" category, we consider a scenario where the attacker has access only to check-ins at venues satisfying that characteristic. 
Each user is characterized by his/her check-in frequency at each location. 
We perform an identification attack on the set of users that check-in at those venues using a Na\"{\i}ve-Bayes attack model~\cite{rossi2014s}. Given a set of check-ins points, a user's identity is estimated with a maximum likelihood approach.
We will refer to the average success rate of the identification attack given a set of venues belonging to a specific type as the \emph{identification complexity} associated with that type, and to a venue type associated with a low identification complexity as \emph{highly discriminative} (and vice versa).

We perform our analysis on 1,391,765 check-ins over 134,989 Foursquare venues distributed in 17 urban regions of the US. The contributions of this work can be summarized as follows:
\begin{itemize}
\item We show that the identification complexity is only weakly correlated with the number of users in a dataset, while it is strongly correlated with the ratio of users to venues. That is, if a large population of users visits a small number of venues, the identification task will be harder, as the check-in patterns of the users tend to overlap.
\item We experimentally demonstrate that the different categories of venues are associated with different identification complexities. Venues in the ``Residence" and ``Travel" category generally are, respectively, the most and least discriminative across most of the urban regions. 
\item We find that, if enough data is available to the attacker, venues in other categories such as ``Shop" become highly discriminative, with an average of 80\% users visiting venues in this category correctly identified.
\item We find that popular venues are less discriminative than unpopular ones, and that spatially isolated venues are significantly more discriminative than venues locate in dense regions. In particular, we show that the 10\% most popular venue are associated with a discriminative power which is almost 3/4 of that associated with the 10\% least popular venues.
\item We show that users with a low entropy of their check-ins distribution are not necessarily the hardest to identify, suggesting that it is the collective behaviour of the user population that determines the complexity of the identification task, rather than the individual behaviour.
\end{itemize}

We believe that the results of this study are of significant interest for individual users of LBSN services, companies managing LBSN data, and governments setting the policies regulating this sector. As far as LBSN users are concerned, our results highlight and quantify potential privacy risks for individuals in relation to the venues they visit.
On the other hand, when releasing a dataset of LBSN check-ins, our analysis can drive the obfuscation of check-ins associated with sensitive venues. We also underline the fact that the analysis proposed in this paper can be applied to any dataset where semantic information is associated to users' venues.
Finally, as already noted in~\cite{rossi2014s}, while it is true that a user implicitly agrees to disclose his/her identity by choosing to participate in a LBSN, a potential attacker can use the LBSN data to transfer the identity information to an external anonymized database~\cite{dwork2008differential}. Note that such a database may contain potentially sensitive data, such as health or financial information. In general, the possibility of linking information across different databases is a well known privacy threat~\cite{sweeney2002k,narayanan2008robust,dwork2008differential,chow2011trajectory,ma2013privacy}. 

The remainder of this paper is organized as follows. We first review the related work, introduce the dataset used in this paper and the necessary data pre-processing. We then review the attack model that we use to assess the identification complexity of a dataset, as well as the classes of venues that on which we focus our analysis. Finally, we present and discuss our experimental findings and we point out potential directions for future research.  

\section{Related Work}

\begin{figure}[t!]
\begin{center}
\includegraphics[width=1\linewidth]{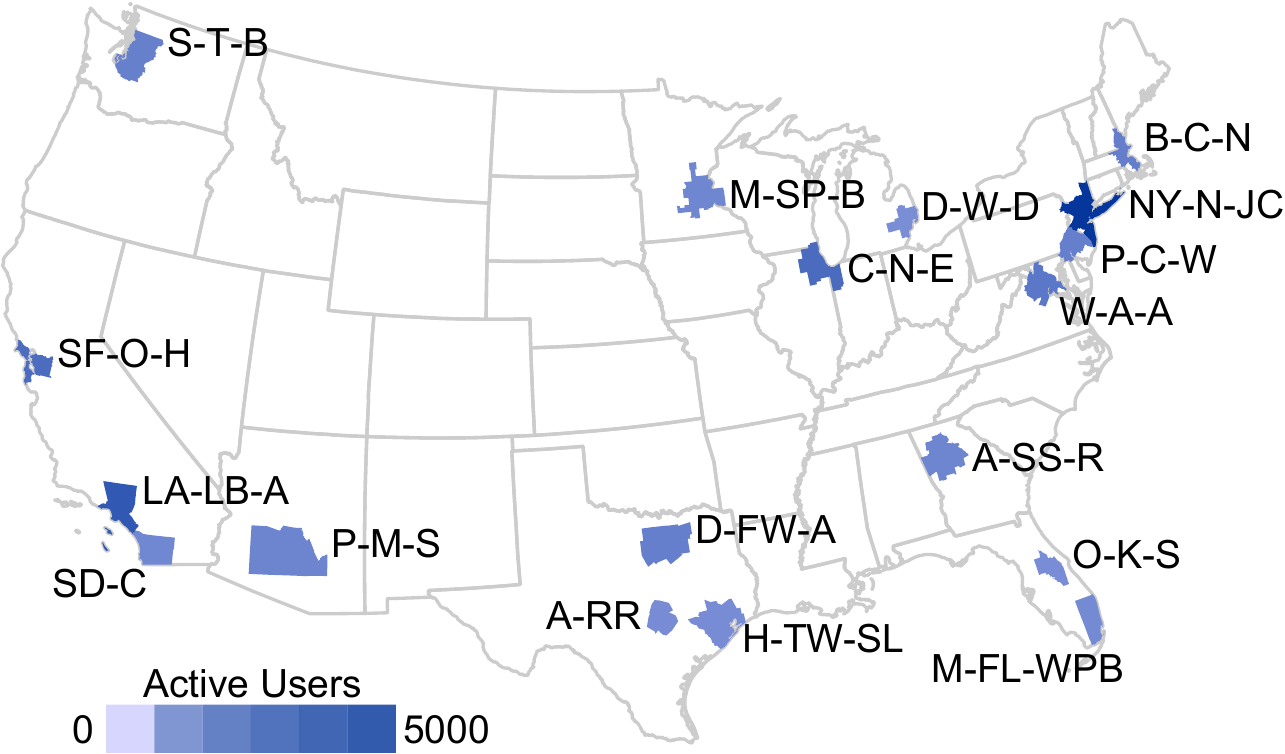}
\caption{Core-based statistical areas (CBSAs) in the US. Only areas used in experimental analysis are included. Areas are shaded by the number of active Foursquare users.}
\label{fig:chore}
\end{center}
\end{figure}


The field of location privacy has been a very active area of research in recent years. The importance of protecting information concerning a person's home location is highlighted for example in~\cite{golle2009anonymity}, where the authors show how data on the home/work pair can be used to carry out inference attacks to reveal the identity of a user from an anonymized GPS trace. On the other hand, Krumm~\cite{krumm2007inference} studies the inverse problem and shows that it is possible to infer the home location of a user participating in a database of GPS traces. More recently, de Montjoye et al. have measured the privacy of users making or receiving mobile phone calls or text messages~\cite{de2013unique}. They find that very few spatio-temporal points from a location trace are needed to uniquely identify the entire trace and thus the individual.

With respect to traditional Location-Based Services (LBS), the additional social dimension of LBSNs works as an incentive for people to share their location data on the social network. Noulas et al. use LBSN data to study the spatio-temporal patterns of users activity~\cite{noulas2011empirical} and build a model human urban mobility in an attempt to predict the next visited location~\cite{noulas2012tale}. Colombo et al. collect a dataset of Foursquare check-ins over the cities of Cardiff and Cambridge in UK, and measure the regularity and predictability of users' check-ins~\cite{colombo2012you}. They find that check-ins are more regular at ``Home" and ``Work" venues, as opposed to ``Outdoors" venues, where check-ins are less predictable.

The concerns and risks associated to location sharing have been investigated from several different angles. For example, in~\cite{ruiz2011location} the authors analyze a series of location privacy issues for LBSN users and describe possible means of protecting privacy. Lindqvist et al.~\cite{lindqvist2011m} conduct a series of interviews and surveys to understand how people manage their privacy on Foursquare. They find that people can choose not to check-in to places for several reasons, for example when the place is perceived as not interesting or particularly sensitive. Cramer et al.~\cite{cramer2011performing} perform a similar analysis and argue that the ability for LBSN users to selectively share their location can help to partially mitigate privacy issues. Jin et al. investigate how users share the addresses and check-ins at residential venues~\cite{jin2012towards}. They find that users are generally aware of their residential privacy, although some of them regularly expose residential check-ins. Even if a user keeps his residential data private, Pontes et al. show that it is possible to infer the user home location with a high accuracy using publicly available information from Foursquare, e.g., the user's ``tips" and ``todos"~\cite{pontes2012we,pontes2012beware}. Moreover, the authors extend their analysis to Google+ and Twitter, where they investigate the possibility of inferring a user home city from the location of his/her friends~\cite{pontes2012beware}. Zhao et al. propose a framework to protect the privacy of LBSN users when their check-ins are stored on the LBSN server~\cite{zhao2013checking}. Finally, Rossi and Musolesi propose a number of identity attacks against LBSN users~\cite{rossi2014s}. More specifically, they propose to match users from an anonymized dataset of check-ins to one where the identity of the users is revealed by either matching their GPS traces or building a probabilistic model of the users' check-in behaviour. 

With respect to the existing research, we are interested in studying how the characteristics of a venue can affect a user's privacy. In other words, our goal is to determine if there is heterogeneity in the discriminative power of venues in a LBSN.

\section{Data}

We perform our analysis on the LBSN dataset collected by Cheng et al.~\cite{cheng2011exploring}. The data was collected by sampling location status updates from the public Twitter feed during the period from September 2010 to January 2011\footnote{http://infolab.tamu.edu/data/} over the entire planet. The original data consists of 22,387,930 check-ins from 224,804 users distributed all over the world, where 53\% of the check-ins are from Foursquare users. Each check-in is labeled with a $userID$, $tweetID$, tweet content, $venueID$, GPS location, and timestamp. In the case of Foursquare tweets, the text field contains a link to the Foursquare venue page, which in turn can be used to download additional attributes describing the venue.

\subsection{Data Processing}
In this subsection we describe the procedure through which we processed the original dataset.

\subsubsection{Mapping Check-ins to Core Based Statistical Areas.}
We first map each check-in to a specific urban region in the US. That is, for each urban region included in the study, we create a separate dataset of users and venues. Without this preprocessing step, the identification task would be trivial, due to the spatial sparsity of the check-ins and the resulting orthogonality of the users check-ins patterns. Moreover, by repeating our analysis over different urban regions we hope to determine if a particular observed effect, i.e., a class of venues being more or less discriminative, is consistent across the different regions or if it depends on the local features of the urban environment.


\begin{table}[!t]
\centering
\small
\renewcommand{\arraystretch}{1}
\setlength{\tabcolsep}{.3em}
\begin{tabular}{c|c|c|c|c}
Main City & Check-ins & Users & Venues & Region Code\\
\hline
Atlanta & 46,860 & 724 & 5,279 & A-SS-R\\
Boston & 56,818 & 918 & 5,860 & B-C-N\\
Chicago & 112,505 & 1,617 & 10,717 & C-N-E\\
Dallas & 66,932 & 886 & 6,268 & D-FW-A\\
Detroit & 37,917 & 515 & 4,139 & D-W-D\\
Houston & 42,094 & 564 & 4,423 & H-TW-SL\\
Los Angeles & 162,570 & 2,513 & 19495 & LA-LB-A\\
Miami & 39,930 & 616 & 5,319 & M-FL-WPB\\
Minneapolis & 46,490 & 696 & 4,600 & M-S.P-B\\
New York & 316,812 & 4,744 & 24,783 & NY-N-JC\\
Orlando & 35,330 & 541 & 2,658 & O-K-S\\
Philadelphia & 63,385 & 895 & 5,415 & P-C-W\\
Phoenix & 55,343 & 744 & 5,978 & P-M-S\\
San Diego & 50,018 & 686 & 5,144 & SD-C\\
San Francisco & 95,823 & 1,560 & 9,587 & SF-O-H\\
Seattle & 61,338 & 868 & 5,897 & S-T-B\\
Washington & 66,874 & 1,141 & 6,675 & W-A-A\\
\hline
\end{tabular}
\caption{Number of check-ins, users, and venues in the selected CBSAs. For each region we also show the name of the main city and the code used to identify the region in Fig.~\ref{fig:chore}.}
\label{table:stats}
\end{table}

We delineate urban regions according to the US Office of Management and Budget (OMB) definition of \textit{Core Based Statistical Area} (CBSA). Specifically, we use the 2013 CBSA standard\footnote{http://www.census.gov/population/metro/}, which specifies $929$ disjoint metropolitan (over 50,000 individuals) and micropolitan (between 10,000 and 50,000 individuals) areas in the USA. A CBSA is defined as a geographic region consisting of an urban core and adjacent dependent areas that have a high degree of social and/or economic integration with the core. A particular CBSA can consist of one or more cities. For example, the most populous CBSA consists of New York, Newark, and Jersey City. For brevity, we abbreviate CBSAs to their initials, e.g., NY-N-JC. The OMB measures core dependence in terms of patterns of commuting between the core and its nearby counties. CBSAs are used by the US Census Bureau for collating population statistics on a federal scale and, therefore, represent a consistent delineation of urban regions across the United States. This is useful as it serves a common definition of urban region applied over a large country. Furthermore, each CBSA captures a wide range of urban land use, from the inner city to the suburbs. Out of the original dataset of 11 million Foursquare check-ins worldwide, we find around 4.9 million check-ins to be within a CBSA in the United States.

\subsubsection{Getting the Venues Attributes.}
In order to obtain the necessary semantics for each location, we query Foursquare for the categories, user count, and check-in count for each venue. Note that, however, the $venueID$s as given in the dataset of~\cite{cheng2011exploring} do not represent valid Foursquare venue IDs. Thus, in order to obtain a mapping from $venueID$s to Foursquare venue IDs, we expand the short URLs contained in the tweets. While this works for the majority of tweets, not all of the short URLs provide valid Foursquare venue IDs. This most likely represents a change in the underlying Foursquare venue database as the original data was collected in 2011. After filtering for non-existing $venueID$s, we are left with 3,336,445 tweets.

\subsubsection{Filtering Inactive Users.}
In this study, we consider only CBSAs having at least $500$ active users, where a user is active if he/she performs at least $20$ check-ins during the period in which the dataset was collected. 
As a result, we are left with 17 CBSAs, as shown in Fig.~\ref{fig:chore}. In total, our dataset is composed of 20,785 users and 1,391,765 check-ins over 134,989 venues. Table~\ref{table:stats} shows some summary statistics for the selected CBSAs.

\section{Methods}
In this section we briefly review the identification attack used in this paper and we describe the venue characteristics taken into consideration for the analysis of the privacy and identification risks. Since the identification attack itself is not a contribution of this work, we refer the reader to~\cite{rossi2014s} for full details. 

\begin{figure*}[t!]
\begin{center}
\subfigure[1 test point (users)]{\includegraphics[width=0.31\linewidth]{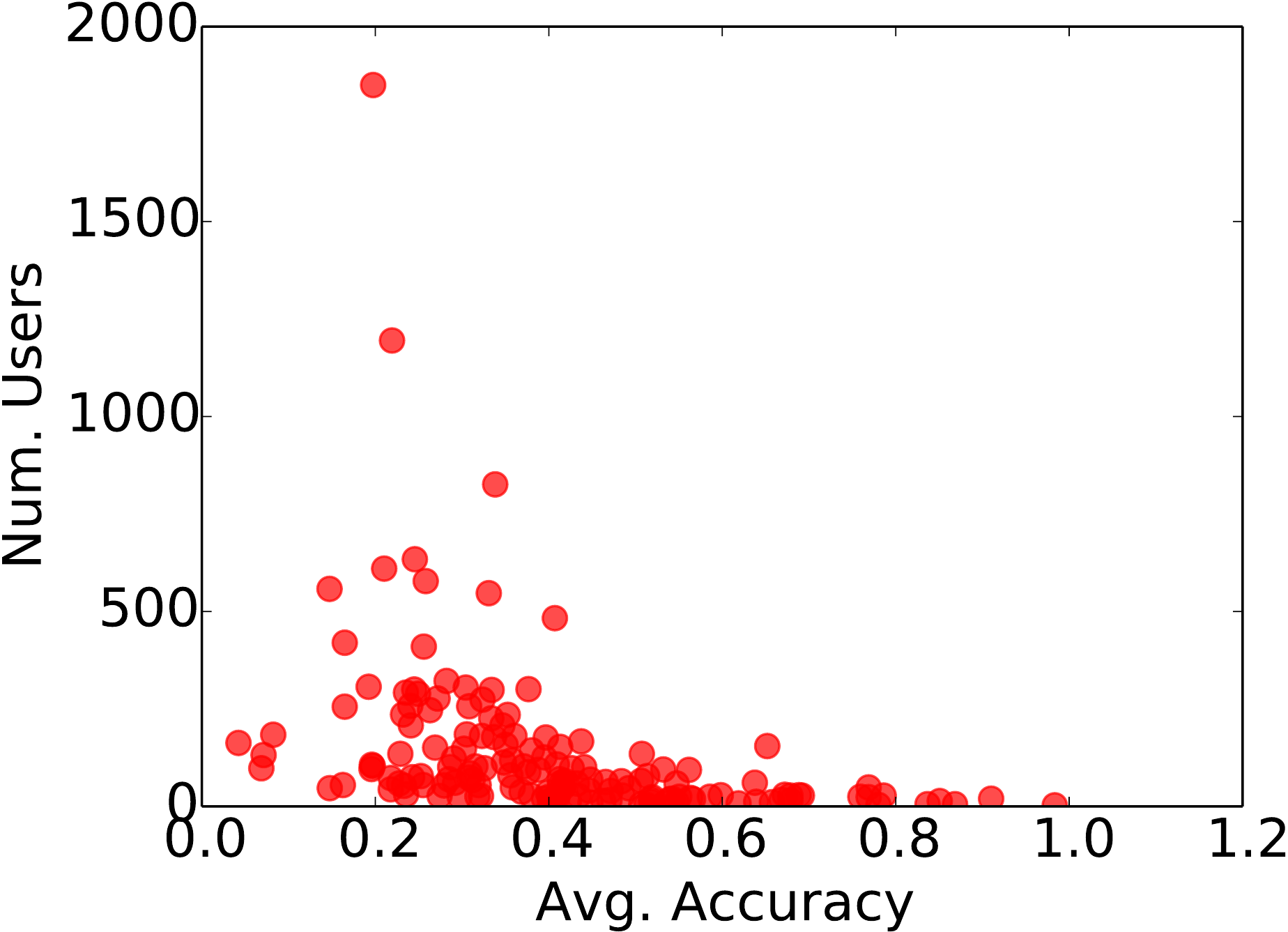}}
\hspace{0.1in}
\subfigure[1 test point (venues)]{\includegraphics[width=0.31\linewidth]{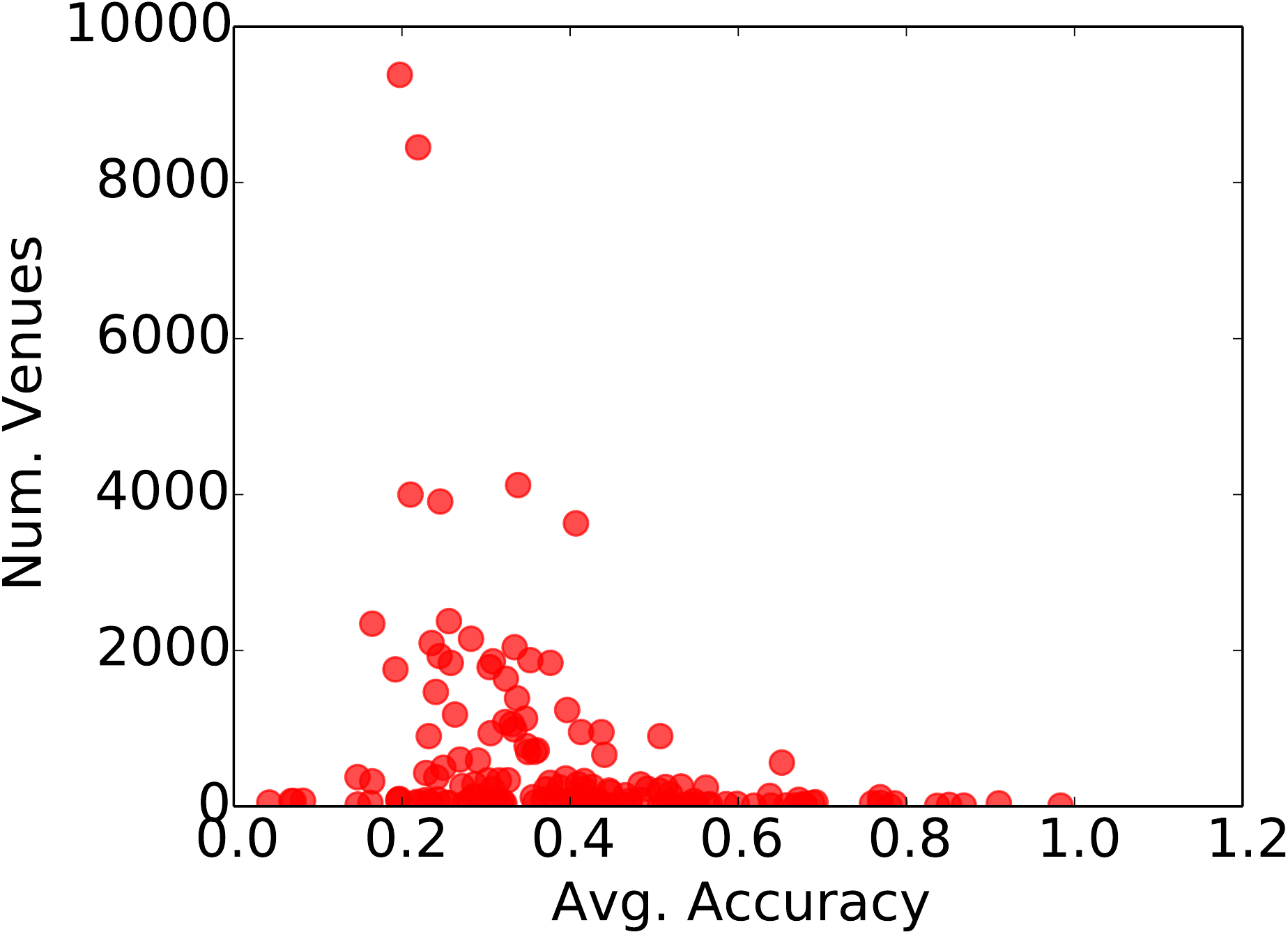}}
\hspace{0.1in}
\subfigure[1 test point (ratio)]{\includegraphics[width=0.31\linewidth]{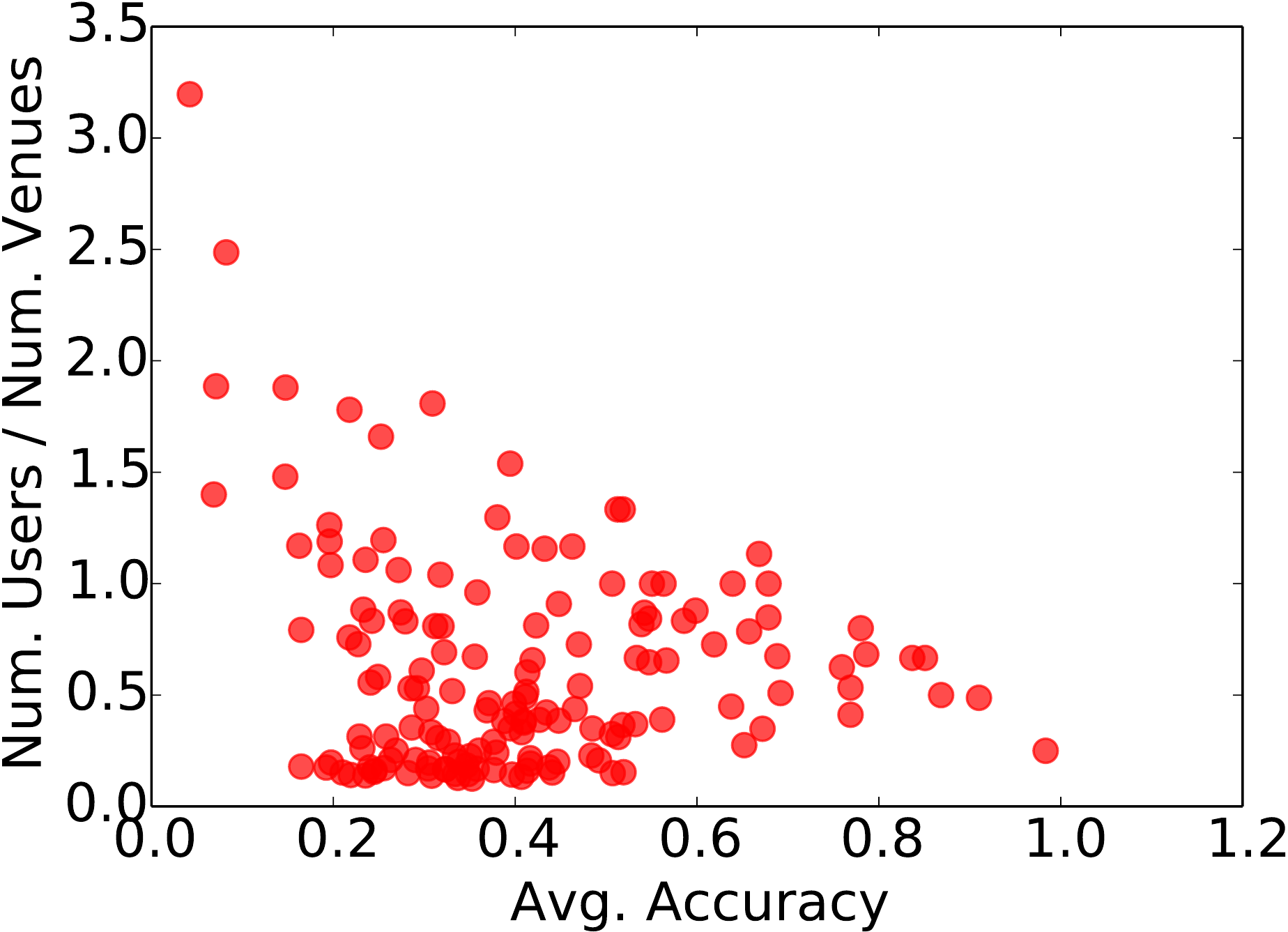}}
\subfigure[10 test points (users)]{\includegraphics[width=0.31\linewidth]{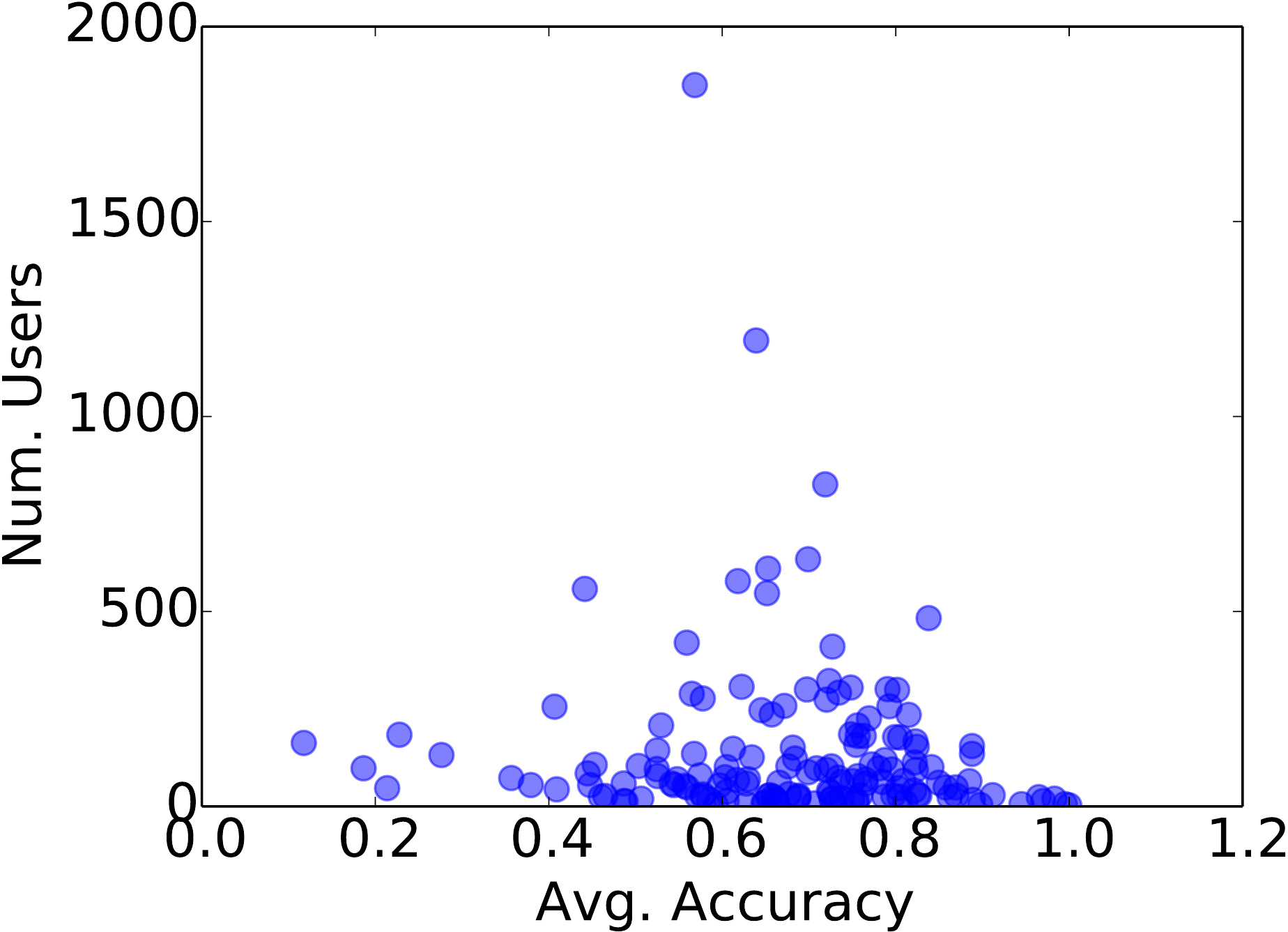}}
\hspace{0.1in}
\subfigure[10 test points (venues)]{\includegraphics[width=0.31\linewidth]{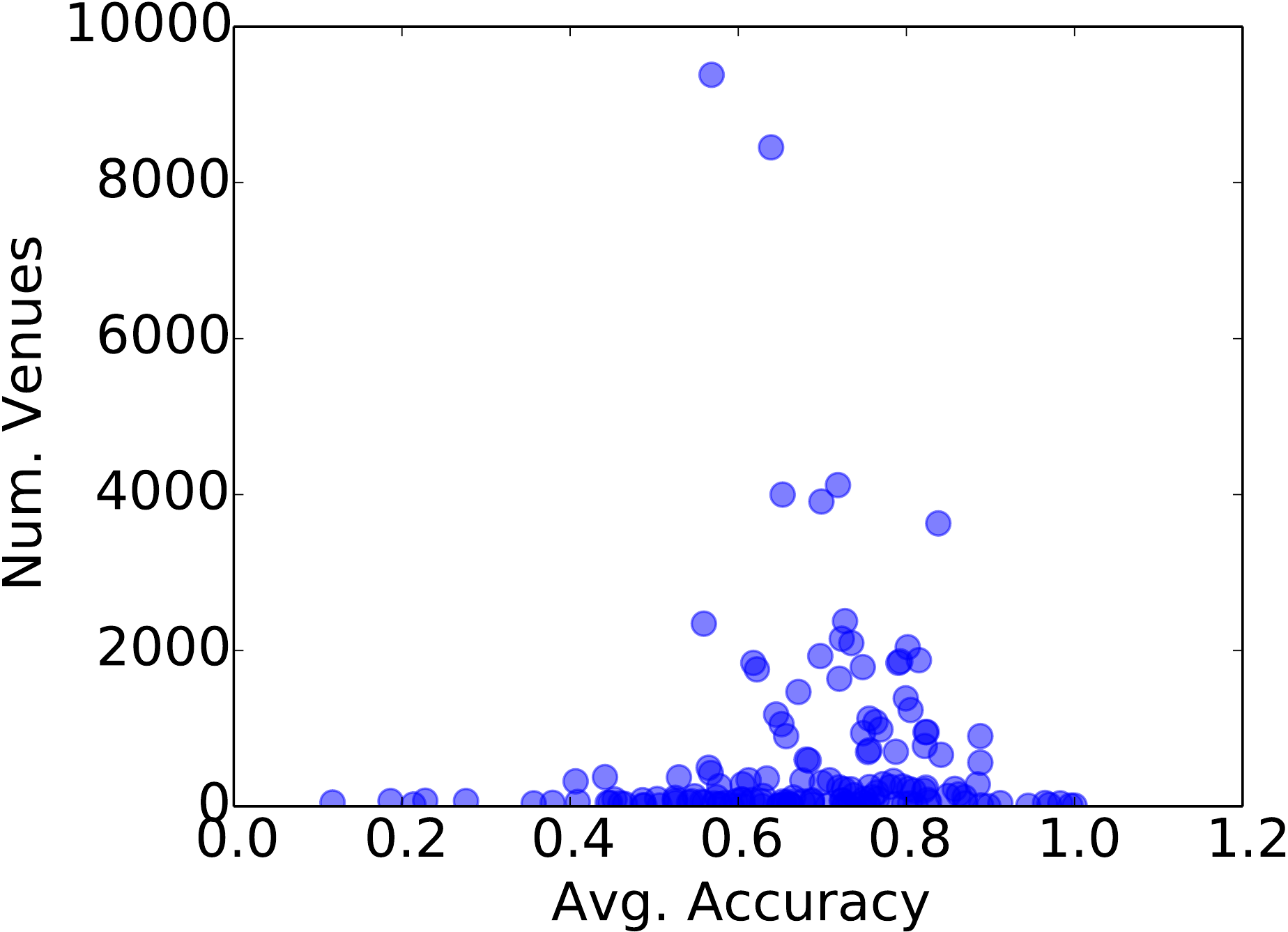}}
\hspace{0.1in}
\subfigure[10 test points (ratio)]{\includegraphics[width=0.31\linewidth]{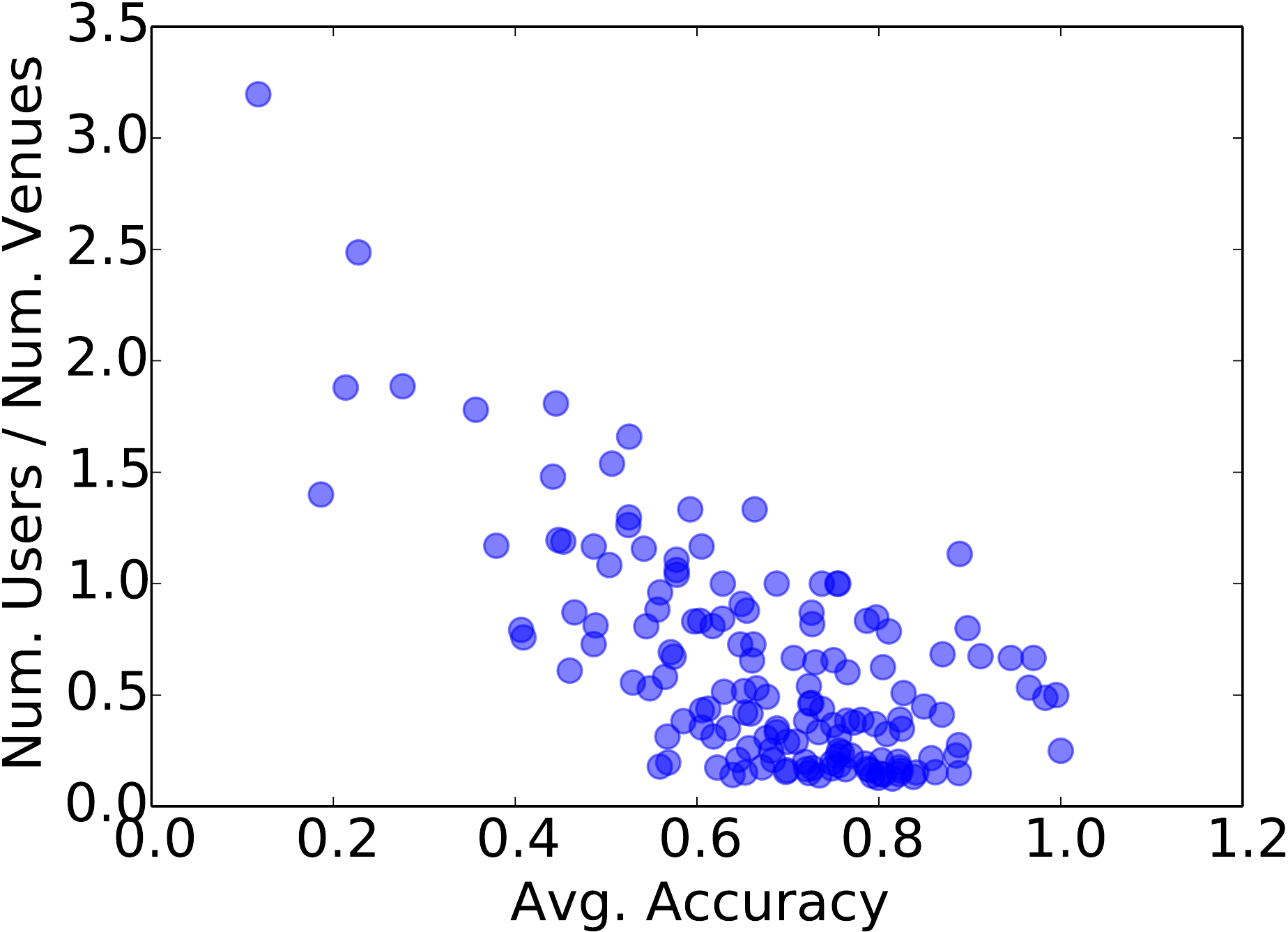}}
\caption{From left to right: scatter plots of the average accuracy versus the number of users, venues, and users over venues ratio, for each combination of region and Foursquare category in the dataset. Here the top and bottom rows show the results for 1 and 10 test points, respectively.}
\label{fig:scatter_avg_accuracy}
\end{center}
\end{figure*}

\subsection{Identification Attack}
Let $C = \lbrace c_1 \ldots c_n \rbrace $ denote a set of check-ins. Each check-in $c_i \in C$ is labeled with a user identifier $u_{id_i}$ and a location identifier $l_{id_i}$. Let $C(u)$ denote the set of check-ins $c_i$ with $u_{id_i} = u $ and $u \in U$, where $U$ is the set of users. Furthermore, let $C(u)$ be divided into a training test $C_{train}(u)$ and a test set $C_{test}(u)$, where in the latter the user identifier is removed. That is, the user $u$ is associated with the set of check-ins $C_{train}(u)$ in the training set, and the set of check-ins $C_{test}(u)$ in the test set. We assume that the attacker has access to a sequence $C_{train}(v)$ for each $v \in U$ in order to identify $u$ from a set of check-ins $C_{test}(u)= \lbrace c_{test_1} \ldots c_{test_m} \rbrace$, i.e., his/her goal is to associate $C_{test}(u)$ to the most probable $C_{train}(v)$.

It is possible to solve this problem by looking for the user $v$ who maximizes the posterior probability
\begin{equation}
v^* = arg\,max_{v \in U} P(v | c_{test_1} \ldots c_{test_m})
\end{equation}
where $P(v | c_{test_1} \ldots c_{test_m})$ denotes the probability of $v \in U$ being the user who generated the check-in series $C_{test}(u)$. This in turn is equivalent to maximizing 
\begin{equation}
v^* = arg\,max_{v \in U} P(v) \prod_{i=1}^m P( c_i | v)
\end{equation}
where $P(v)$ is the user prior and $P(c_i | v)$ is the probability of $c_i$ being a check-in generated by $v$. The attack will be successful when $v^* = u$. Under the assumption of a uniform distribution for the user prior, the multinomial distribution associated to each user can be estimated using a standard maximum likelihood approach, i.e.,
\begin{equation}
P(c_i | v) = \frac{N^v_i + \alpha}{\sum_{j=1}^n N^v_j + \alpha |L|} 
\end{equation}
where $N^v_i$ denotes the number check-ins of $v$ at the location $l_{id_i}$ in $C_{train}(v)$, $n$ is the number of locations visited by $v$, $|L|$ is the number of locations in our dataset and $\alpha>0$ is a smoothing parameter to remove zero-probabilities~\cite{manning2008introduction}.

\subsection{Venue Characteristics}
For each venue in our dataset, we obtain the following set of attributes: 1) venue category; 2) number of (unique) visitors; 3) number of visits; and 4) distance from the nearest venue. Here \emph{category} refers to any of the top level elements in the Foursquare category hierarchy\footnote{https://developer.foursquare.com/categorytree}.  We do not consider venues in the ``Event" category as after processing the original dataset we found that some CBSAs do not have any check-in at venues of type ``Event". The \emph{number of visitors} and the \emph{number of visits} measure the popularity of a venue, where the latter can be also seen as a proxy for the entropy of a venue, i.e., a measure of the diversity of unique visitors
to a location. Thus, it is reasonable to expect very popular venues and high entropy venues to be associated with a higher identification complexity, as the they are characterized by a higher overlap of the users' check-in patterns. Finally, the \emph{distance from the nearest venue} is used as a measure of the spatial isolation of a venue. This spatial feature captures the property that venues are not distributed evenly throughout urban regions. Urban regions typically feature a small number of highly dense areas, such as business districts and retail centers, whereas venues are more sparse at the peripheries of a city.

\begin{figure*}[t!]
\begin{center}
\includegraphics[width=1\linewidth]{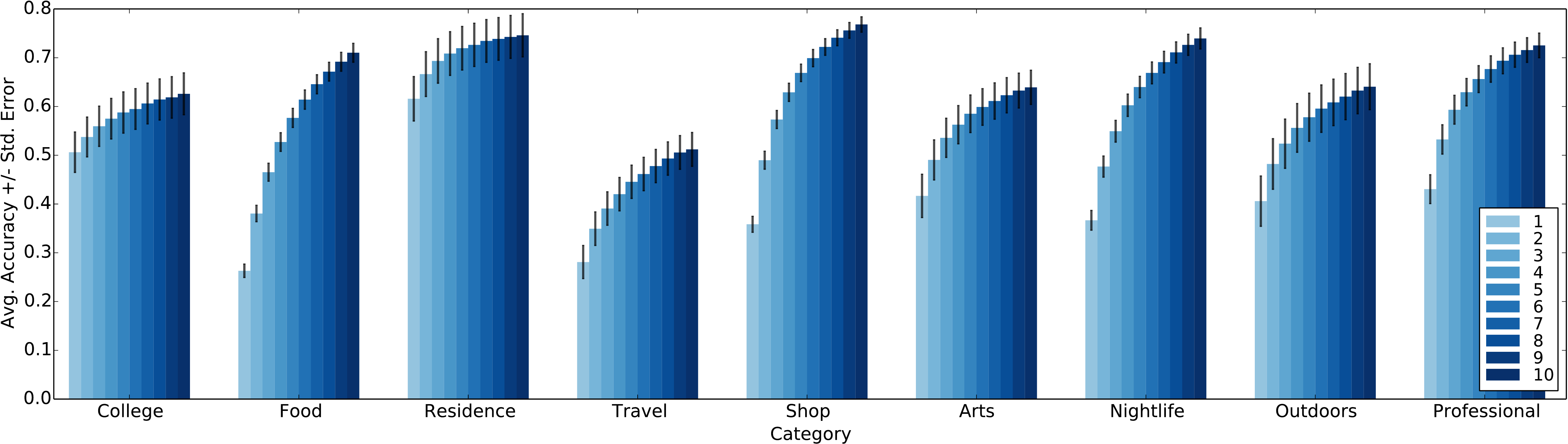}
\caption{Average classification accuracy for each category over all the regions, for increasing number of test points (1 to 10). When the test set contains only a few check-ins the most discriminative is the ``Residence". Venues in the ``Travel" category, which are generally characterized by a high user to venues ratio, correspond to a relatively low identification complexity.}
\label{fig:ranking}
\end{center}
\end{figure*}

Note that while the \emph{category} attribute is discrete, the other attributes are continuous-valued. Given a continuous-valued attribute, e.g., number of visitors, we perform a percentile-based discretization. That is, we rank the possible values of the continuous attribute $a$ and we partition them into equally-sized groups. Then, each venue is assigned to a \emph{class} based on the value of its attribute. In the remainder of the paper, we will refer to a ``class of venues" as a group of venues characterized by a common attribute, i.e., venues in the ``Food" category, the ``10\% most popular" venues, the ``20\% least isolated venues".

\section{Analysis}

Recall that our goal is to determine which classes of venues are correlated with a high identification complexity. In other words, what are the venues that a malicious user should monitor to increase the probability of success of the identification attack? To answer this question, we simulate a scenario in which, given a class of venues, the attacker has access only to the check-ins that belong to that class.

\subsection{Experimental Setup}
In order to understand the relation between the specific types of venues and complexity of the identification task, let us assume that the attacker has access only to a number of check-ins in locations in specific categories.
For example, let us assume that an attacker will be able to access only 
``Food" venues, e.g., restaurants. Given this \emph{training} data, the attacker is then able to learn a model of check-in frequencies of users over the venues belonging to the ``Food" category.

We then suppose that the attacker has access to a source of anonymized location information, i.e., a \emph{test} set. This could be both in the form of LBSNs check-in data or sequences of GPS points, where the latter can be reduced to a finite set of venues by extracting the set of significant places~\cite{ashbrook2003using} and mapping them to Foursquare venues. The attacker's aim is to reveal the identities of the participants by linking the location information in the training set to the anonymized test set. Note that this scenario is essentially equivalent to one in which the attacker trains the model on all the check-ins, without any restriction on the venue class, and tests it on a subset of the data. In fact,
it is highly unlikely that a user that has been observed at a certain venue in the test set will be matched to a user that was never at that venue in the training set\footnote{More precisely, such an event will be associated with a small but non-zero probability. This is a consequence of the fact that the Bayesian model used for identification assigns non-zero probabilities to the presence in all the venues.}.
%

%

In order to simulate the scenario described above, we consider check-ins belonging to specific categories separately.
More specifically, for each user $u$ we separate his/her check-ins into a training set $C_{train}(u)$ and a test set $C_{test}(u)$.
We measure the performance of the identification attack in terms of classification accuracy, i.e., the ratio of successfully identified users. For each region and each venue class, we repeat the identification attack 100 times to compute the average classification accuracy ($\pm$ standard error). Finally, note that for each user we keep the training set size fixed, while we vary the size of the test set $C_{test}(u)$ from 1 to 10 check-ins. That is, we measure the classification accuracy of the attack as the number of anonymized samples observed by the attacker varies from a minimum of 1 check-in to a maximum of 10 check-ins.

Recall that, before running the identification attack, we remove all the check-ins at venues that do not belong to the type being investigated. According to this experimental setup, the maximum number of users that the attacker can successfully identify varies according to the number of active users that are left after the filtering. In other words, the inherent complexity of distinguishing among $k$ users is higher as $k$ increases, i.e., a random guess will return the correct identity of a user with probability $\frac{1}{k}$. We believe that this does not hinder our analysis. First, the fact that only a small set of users is actively visiting a certain class of venues is a confirmation that that class of venues is highly discriminative in itself. Second, here we assume that the aim of the attacker is to maximize the success of his/her attack rather than the number of the identified users. This could be, for example, because the cost of a misclassification is very high. Finally, whereas the lower bound on the identification complexity is clearly negatively correlated with the number of users, our experimental results show only a weak linear correlation between the latter and the identification complexity.

\subsection{Categories and Identification Complexity}
\begin{figure*}[t!]
\begin{center}
\subfigure[1 test point]{\includegraphics[width=0.48\linewidth]{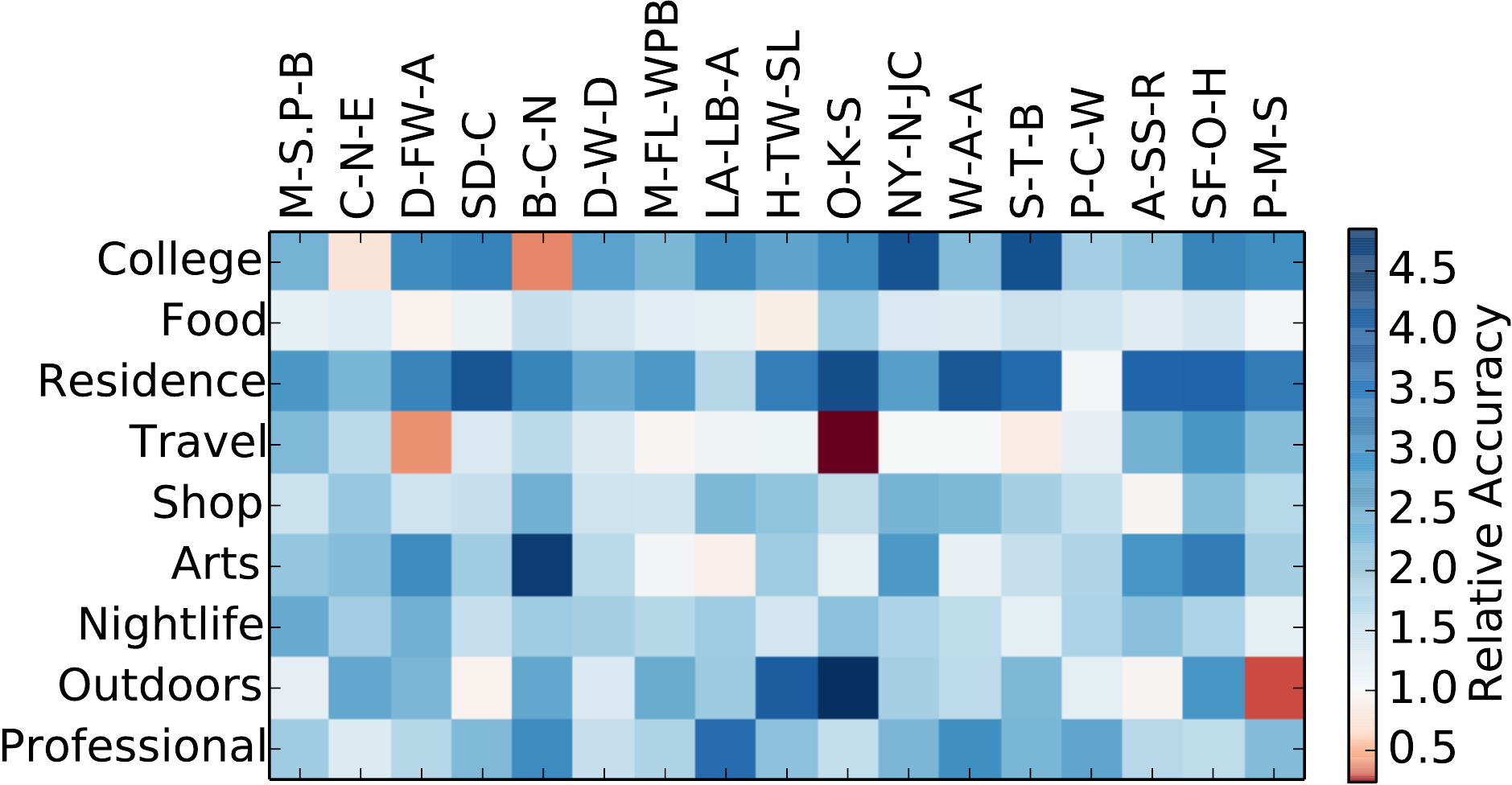}\label{fig:rankings_heatmap_1}}
\hspace{0.1in}
\subfigure[10 test points]{\includegraphics[width=0.48\linewidth]{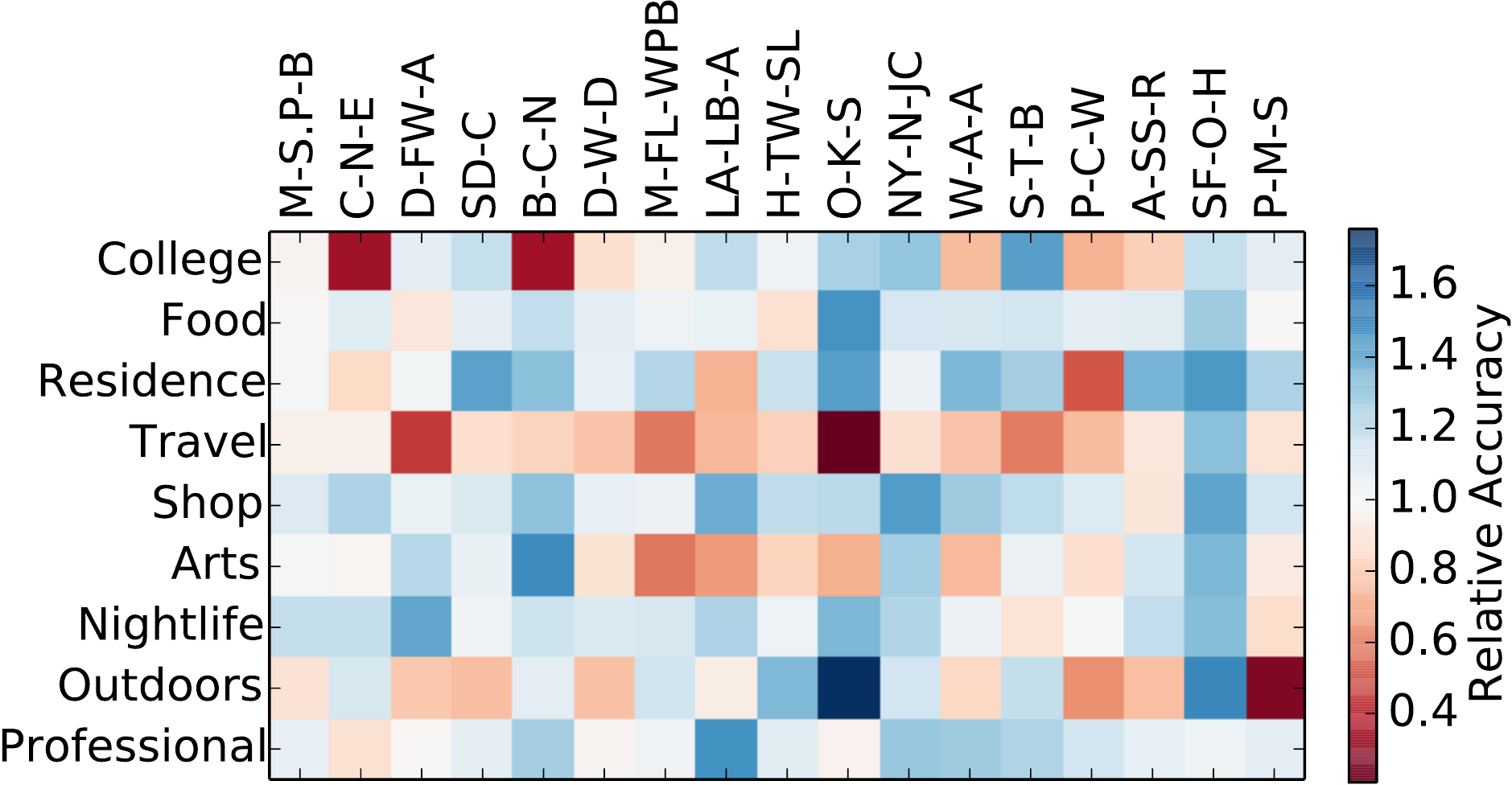}\label{fig:rankings_heatmap_10}}
\caption{Relative classification accuracy for each region and category with 1 (a) and 10 (b) test points (best viewed in color). The relative accuracy is computed by dividing the average accuracy for a given region and category by accuracy in that region when all categories are considered. Thus, it can be interpreted as the relative gain (in terms of identification success) that the attacker gets when narrowing his/her attack on a specific venue category.}
\label{fig:rankings_heatmap}
\end{center}
\end{figure*}

As a first experiment, we measure how the identification complexity varies when the attacker observes check-ins from different Foursquare categories. Fig.~\ref{fig:scatter_avg_accuracy} shows the average classification accuracy versus the number of users, venues and user over venues ratio, for each region and venue class. In other words, each point corresponds to a different identification scenario. The top row shows the average accuracy for $|C_{test}(u)| = 1$, while the bottom row shows the average accuracy for $|C_{test}(u)| = 10$. We compute the Pearson correlation coefficient for all the six scatter plots and we find that when the test size is 1, i.e., the attacker observes a single check-in from an unidentified user, there is linear correlation of $-0.427$ (p-value $< 0.01$) between the average accuracy and the number of users. We also find a linear correlation of $-0.292$ (p-value $< 0.01$) between the average accuracy and the number of venues. On the other hand, if the attacker observes each anonymous users performing 10 different check-ins, the correlation with the number of users as well as the correlation with the number of venues drops to zero, while we measure a linear correlation of $-0.685$ (p-value $< 0.01$) with the ratio of user over venues. In other words, when a large population of users visit a limited number of venues, their check-in patterns are likely to be more similar, and then it is more difficult for the attacker to discriminate among the anonymous users. On the other hand, when only 1 observation is available to the attacker, his/her guess will be close to a random one, and, as we observed earlier, the accuracy of a random guess is inversely proportional to the number of users in the problem.

Fig.~\ref{fig:ranking} shows the average classification accuracy over the 17 CBSAs, for each Foursquare category and for increasing number of test points. When the number of test points is limited, venues in the ``Residence" class are the most discriminative ones, followed by venues in the ``College" class. However, as the test set size increases, the ``Residence" class remains highly discriminative, while ``College" venues drop among the least discriminative, and other types of venues such as ``Shop", ``Food" and ``Nightlife" rise among the most discriminative. Note that Foursquare users are already advised to avoid disclosing information about residential locations, in particular venues in the ``Home" category\footnote{https://foursquare.com/privacy}. However, our results show that as little as 10 points are sufficient to identify a user with a high confidence, even if the points belong to a different class. Interestingly, Fig.~\ref{fig:ranking} also shows that the least discriminative venues are in the ``Travel" category. Intuitively, venues belonging to this class, such as railway stations and airports, are characterized by a higher entropy, i.e., a higher user to venues ratio, and thus lead to a higher identification complexity.  

\subsection{To Filter or not to Filter}

\begin{figure*}[t!]
\begin{center}
\subfigure[Average Accuracy]{\includegraphics[width=.4\linewidth]{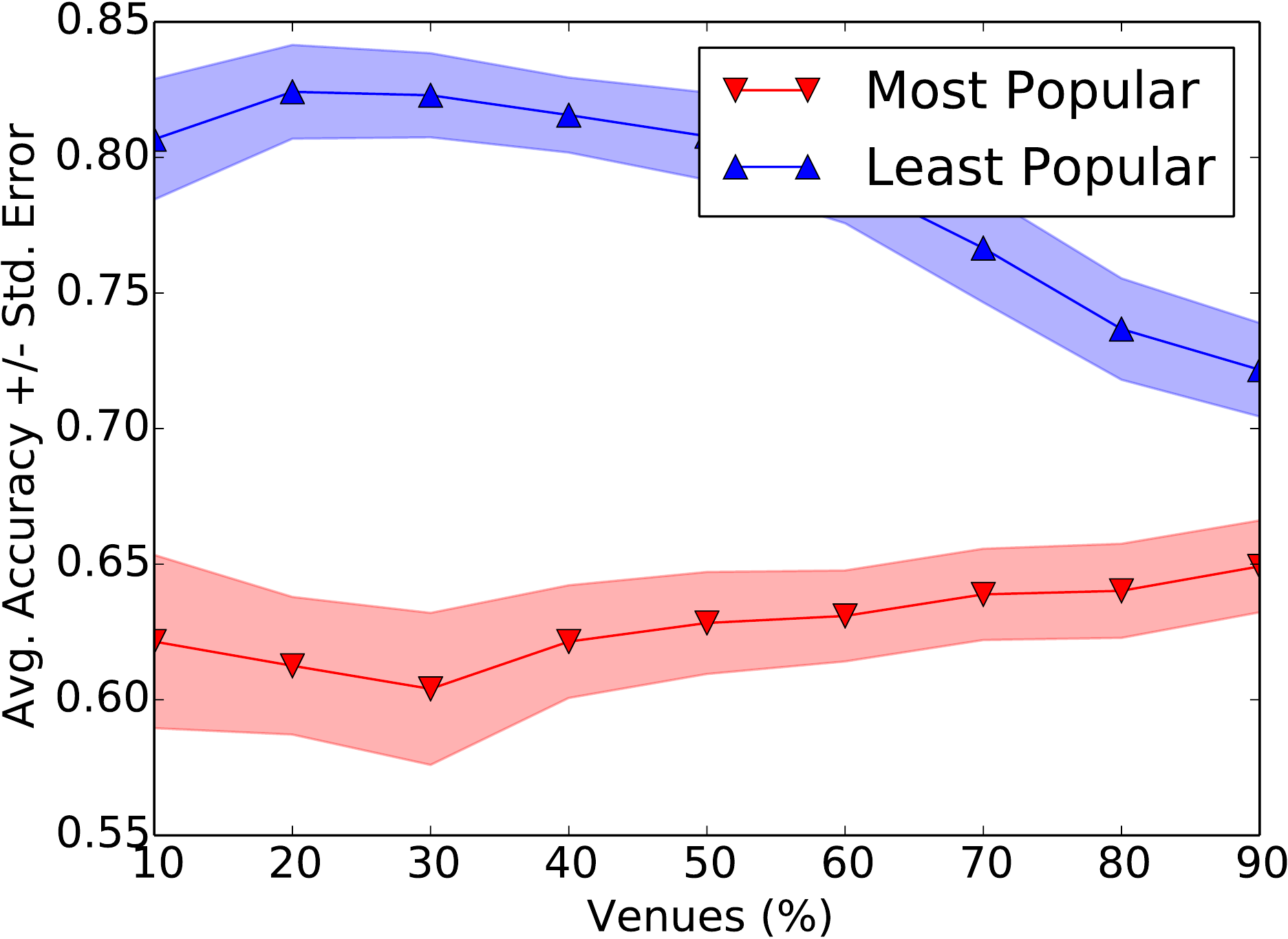}\label{fig:venue_popularity}}
\hspace{0.5in}
\subfigure[Average Number of Users]{\includegraphics[width=.4\linewidth]{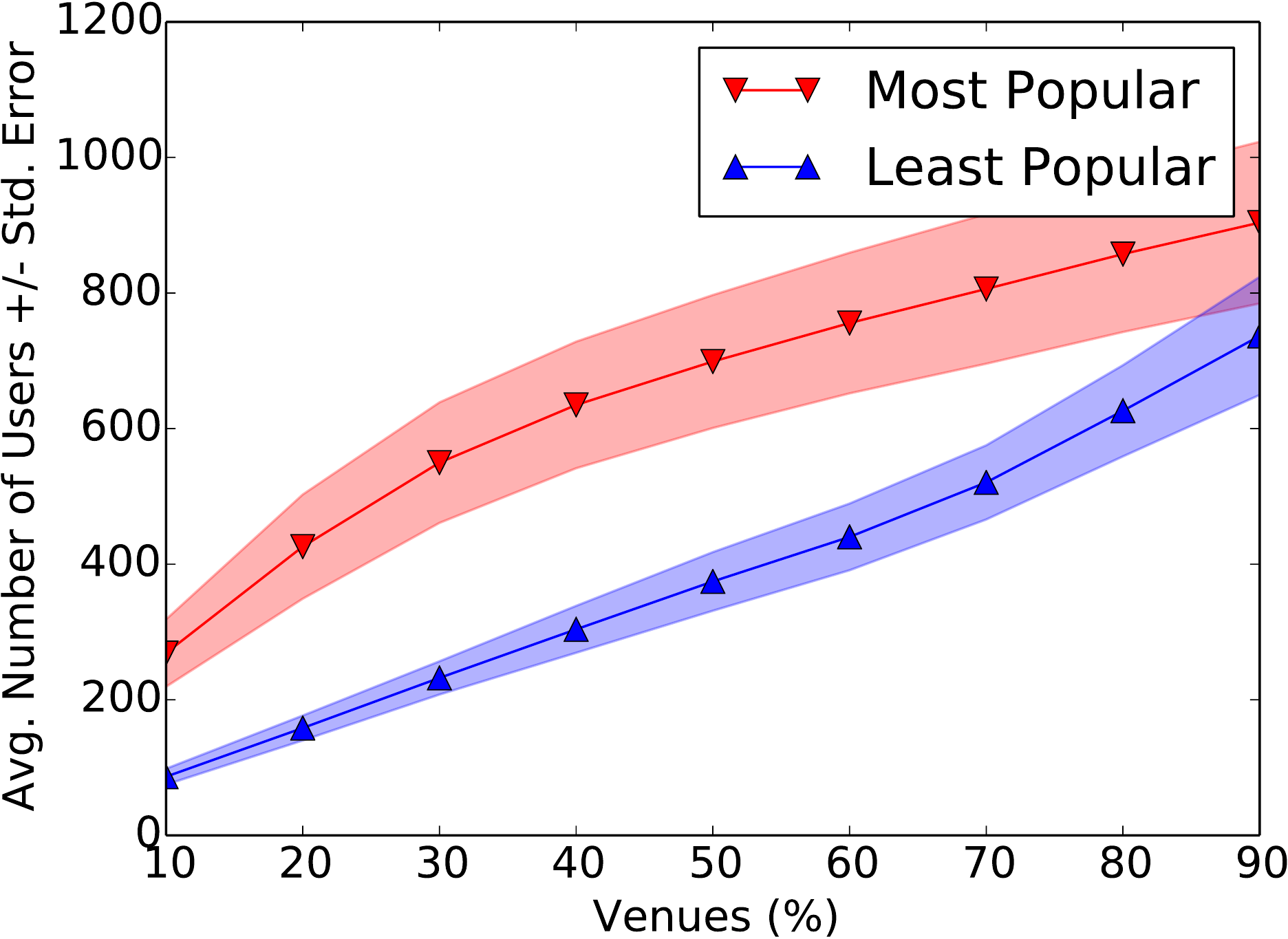}\label{fig:users_venue_popularity}}
\caption{Average accuracy and average number of active users for increasing percentiles of top/least popular venues. Note that as we include increasing percentiles of venues the number of users to be identified grows, and it is generally higher for more popular venues. Here the average accuracy is computed for a test size of 10.}
\end{center}
\end{figure*}

\begin{figure*}[t!]
\begin{center}
\subfigure[Average Accuracy]{\includegraphics[width=.4\linewidth]{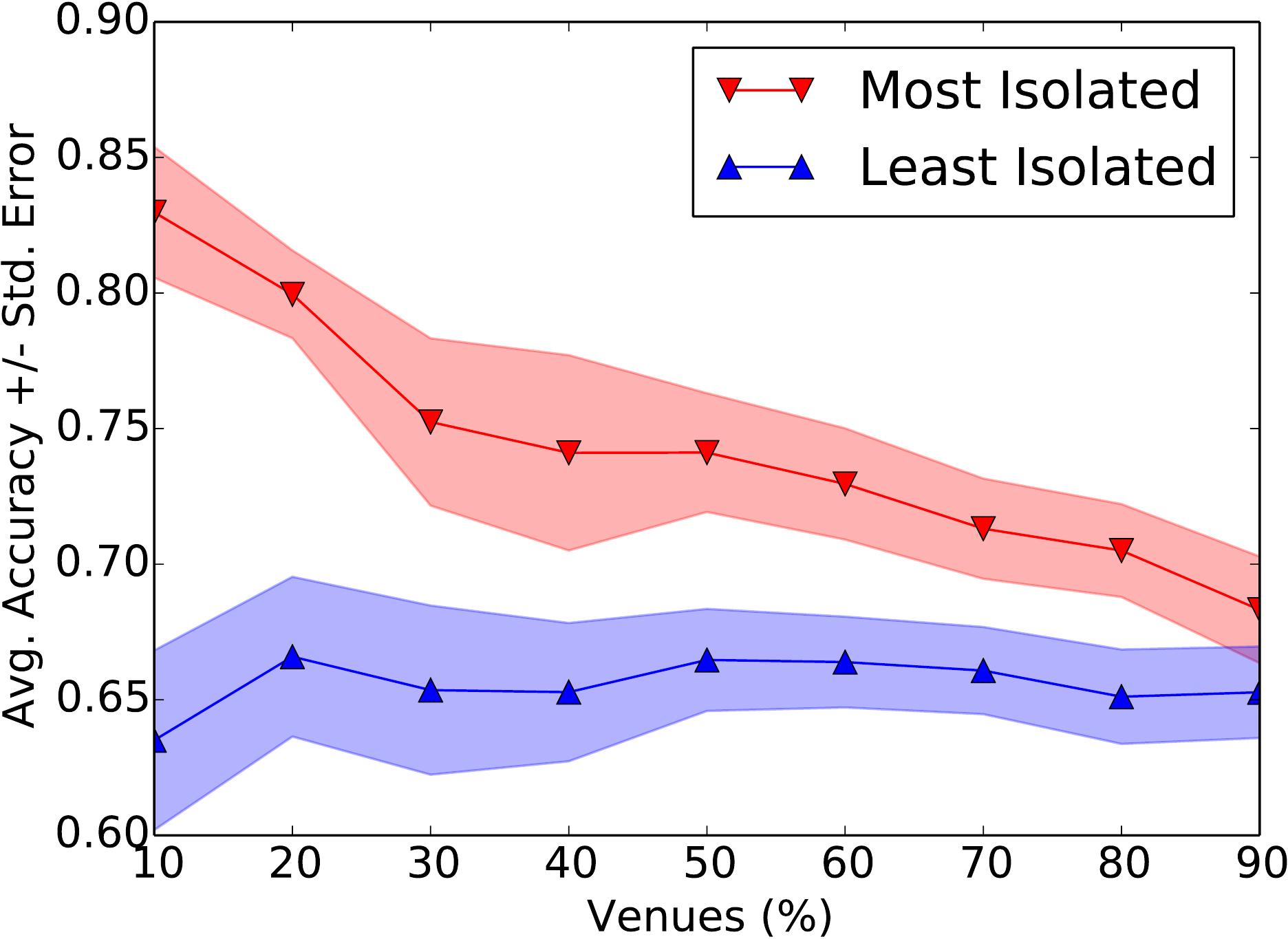}\label{fig:venue_location}}
\hspace{0.5in}
\subfigure[Average Number of Users]{\includegraphics[width=.4\linewidth]{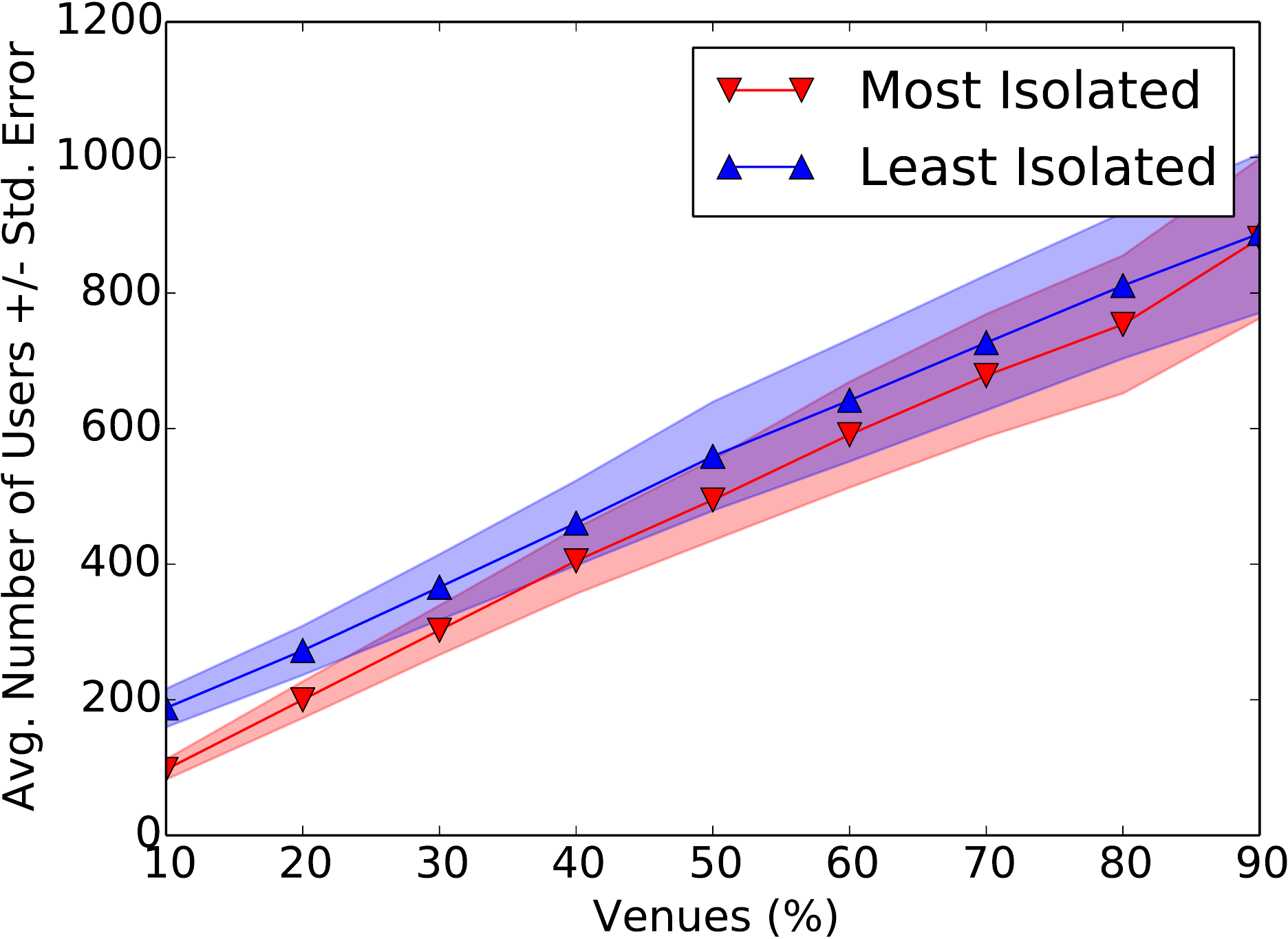}\label{fig:users_venue_location}}
\caption{Average accuracy and average number of active users for increasing percentiles of top/least isolated venues. Here the average accuracy is computed for a test size of 10. Note that given a certain proportion of venues the number of users in the classification problem is close or equal, regardless of the fact that we are considering popular or unpopular venues. }
\end{center}
\end{figure*}

So far, we have focused our attention on the classes of venues that yield the highest identification accuracy, but we have not considered what the result would have been by building our user models using all the data available, i.e., the check-ins over all the venue categories. 

In other words, if the attacker had access to the whole dataset of check-ins, would it be really more effective to focus on a specific class of venues (i.e., consider only venues belonging to certain categories)? To answer this question, we compute again the average classification accuracy for each of the 17 CBSAs. Note that now we only remove inactive users, but we do not apply any filtering to the venue categories. For each region and category, we compute the relative accuracy by dividing the corresponding classification accuracy by the average accuracy over the entire set of check-ins of the region.

Figs.~\ref{fig:rankings_heatmap_1} and~\ref{fig:rankings_heatmap_10} show the relative average accuracy over each region and category, when the test size is 1 and 10, respectively. Note that the white color is used to indicate a relative accuracy of 1, i.e., no gain or loss with respect to the attack that makes use of all the data.
For example, the relative accuracy of the ``Food" category in San Diego-Carlsbad (SD-C) is a measure of how discriminative the ``Food" category is in SD-C, compared to the accuracy that can be achieved using all the available data for that region. Fig.~\ref{fig:rankings_heatmap_1} shows again that when the test set contains as little as 1 check-in, the ``Food" category is the among the least discriminative across all regions, while the ``Residence" category is among the most discriminative. On the other hand, the performance for the different classes are similar when the test set size is 10. Note again that when the attacker observes a single anonymized check-in, the limited information available implies that his/her guess on the the identity of the anonymous will be close to a random guess. This in turn explains why when the test size is 1, the relative accuracy is usually greater than 1. On the other hand, as we increase the size of the test set, the attacker's confidence on his/her prediction also grows, and we find that check-ins in the ``Travel" category are relatively poor predictors of a person's identity.

Finally, Figs.~\ref{fig:rankings_heatmap_1} and~\ref{fig:rankings_heatmap_10} provide an insight into the differences and similarities among the various regions, as well as the ranking of the different categories at city level. For example, while the ``College" category is usually highly ranked in terms of average accuracy, in the CBSA of Boston-Cambridge-Newton (B-C-N) this corresponds to the least discriminative category. This may be due to the fact that the region of Boston hosts a considerable number of notable colleges and universities, as well as large student population. In fact, for this category of venues we observe the highest users to venue ratio in the region of Boston. 


\subsection{Influence of Venue Popularity}

In this subsection we analyze the identification complexity associated with venues of varying popularity. More specifically, we measure the popularity of a venue in terms of the number of unique users that visited that venue. Intuitively, we expect less popular locations, i.e, a niche coffee shop or a residence, to be more discriminative than popular locations, i.e., a train station. Fig.~\ref{fig:venue_popularity} shows how the average classification accuracy varies as we take an increasing proportion of the top and least popular venues, where the accuracy is averaged over all the CBSAs. As expected, we find that it is harder to identify users that check-in at very popular venues. It may be tempting to explain this effect by conjecturing that the lower accuracy over popular venues is a result a higher number of users. Fig.~\ref{fig:users_venue_popularity} shows that indeed the most popular venues are (by definition) visited by a larger population of users. However, even if we choose two proportions of top and least popular venues such that the number of active users is roughly the same in both sets, we still observe a higher accuracy in the case of the least popular venues. In fact, the observed effect can be explained by the higher entropy of the most popular venues. Most notably, in the case of the least popular venues, as we start including venues with a higher users' count the accuracy starts to drop. This is indeed a consequence of both the higher number of users, as well as the fact that we are adding high entropy venues to the problem. Note again that this is not merely a result of the larger set of users. In fact, in the case of the most popular venues, as we increase the number of venues in our dataset to include the least popular ones, the accuracy increases or remains stable. This is in strong contrast with the intuition that adding more users to the identification problem makes it harder.

\begin{figure}[t!]
\begin{center}
\includegraphics[width=.95\linewidth]{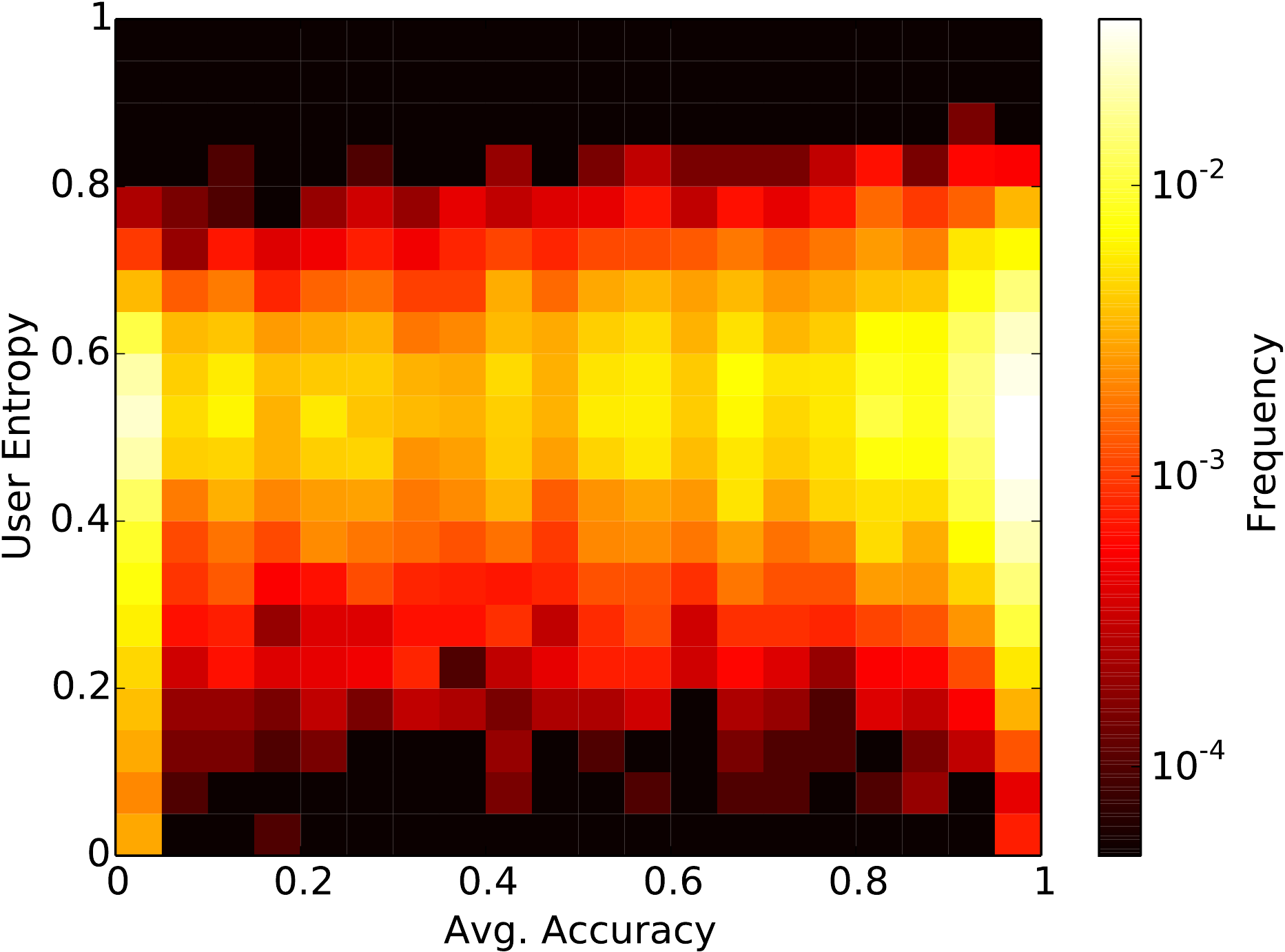}
\caption{Heat map showing the relation between users entropy and identification accuracy, where the user entropy is the Shannon entropy of the histogram of the user's check-ins. Here a brighter color indicates a higher concentration of points. Note that the average accuracy is computed per user, and it corresponds to the proportion of successful attacks against a user's identity.}
\label{fig:entropy}
\end{center}
\end{figure}

\subsection{Influence of Venue Location}

We now turn our attention to the influence of a venue geographic position on the complexity of the identification task. More specifically, we measure the impact of the spatial isolation of a venue on the identification accuracy. Fig.~\ref{fig:venue_location} shows how the average accuracy varies as we include increasingly isolated venues in our dataset. We observe a significant difference between the identification accuracy over the most/least isolated venues. In other words, our results show that more accessible venues are less discriminative. Intuitively, venues located in high density areas allow to easily access a large number of alternative venues. This in turn results in a higher venue entropy, and thus an increased identification complexity. Finally, note that the difference in accuracy does not follow from a difference in the number of users, as Fig.~\ref{fig:users_venue_location} shows.

\subsection{Users' Entropy and Identification Task}

As a last experiment, we investigate how the entropy of the distribution of a user's check-ins influences the ability to identify him/her. Intuitively, we expect high entropy users, i.e., users that check-in frequently at a large number of venues, to be less identifiable than users with low entropy, i.e., users that check-in frequently at a few venues. Fig.~\ref{fig:entropy} shows a heatmap of the users distribution, where for each users we compute his/her identification accuracy, i.e., the proportion of successful identity attack that targeted him/her, as well as his/her entropy. Surprisingly, we find that there is no correlation between these two quantities, suggesting that it is the collective behaviour rather than individual behaviour that determines the identification complexity of the individual.

\section{Discussion and Future Work}
In this paper we have investigated the interaction between venues characteristics and users' privacy in LBSNs. We have analyzed over 1 million Foursquare check-ins from 17 urban regions in the US. 
Our experimental analysis has shown that different classes of venues are characterized by different levels of user identity discriminative power. We found that the identification complexity is strongly correlated with the ratio of users to venues, and that venues in the ``Residence" and ``Travel" category are respectively the most and least discriminative across most of the urban regions. Interestingly, we found that venue categories that are not commonly associated with a high identity privacy risk can still be highly discriminative. For example, we found that we can correctly identify 80\% of the users visiting venues in the ``Shop" category. Our results also showed that the popularity of a venue and its spatial isolation are positively and negatively correlated with their discriminative power, respectively. For example, by considering only check-ins at the 10\% most popular venues rather than the 10\% least popular venues, we observed a drop in the classification accuracy from 80\% to 62\%. Finally, we found that there is no correlation between the entropy of a user's check-in frequency and the ability to successfully identify him/her. This in turn suggests that the collective behaviour of the population rather than the individual behaviour has to be taken into account in order to estimate the risk of being identified from location data.

We believe that our findings raise important privacy concerns, but, at the same time, they shed light on potential ways to address these issues. For example, our results are a reminder for LBSN users that check-ins at highly discriminative venues, such as spatially isolated or niche ones, should receive a particular attention by users in terms of public disclosure. 
Our analysis also suggests that it is not how a user distributes his/her check-ins over the venues, i.e., frequently visiting a limited number of venues rather than distributing his/her check-ins over a larger set, that determines how difficult it is to identify him/her. Rather, it is the type of venues that a user visits that matters. As far as the release of anonymized location datasets is concerned, our findings can be used as an indicator of which subsets of the data should potentially be selected when applying obfuscation and anonymization methods. Finally, designers of LBSNs may wish to consider the discriminatory power of categories when implementing their privacy policies.

Future work should investigate the similarities and differences between users' identifiability across different cities and countries. More specifically, we aim to study to what extent the urban environment plays a part in shaping the users check-in patterns and thus their identity privacy. Our results indicate that there is some diversity in the identifiability of users in the cities studied. The reasons for this heterogeneity should be investigated further, through consideration of a wider variety of cities and countries, as well as LBSNs platforms.

\section*{Acknowledgement}
This work was supported through the ``The Uncertainty of Identity: Linking Spatiotemporal Information Between Virtual and Real Worlds'' Project (EP/ J005266/1), funded by the EPSRC, and the ``LASAGNE'' Project, Contract No. 318132 (STREP), funded by the European Commission.

\bibliographystyle{aaai}
\bibliography{biblio}

\begin{thebibliography}{}

\bibitem[\protect\citeauthoryear{Ashbrook and
  Starner}{2003}]{ashbrook2003using}
Ashbrook, D., and Starner, T.
\newblock 2003.
\newblock {Using GPS to learn significant locations and predict movement across
  multiple users}.
\newblock {\em Personal and Ubiquitous Computing} 7(5):275--286.

\bibitem[\protect\citeauthoryear{Beresford and
  Stajano}{2003}]{beresford2003location}
Beresford, A.~R., and Stajano, F.
\newblock 2003.
\newblock Location privacy in pervasive computing.
\newblock {\em IEEE Pervasive Computing} 2(1):46--55.

\bibitem[\protect\citeauthoryear{Bettini, Wang, and
  Jajodia}{2005}]{bettini2005protecting}
Bettini, C.; Wang, X.~S.; and Jajodia, S.
\newblock 2005.
\newblock Protecting privacy against location-based personal identification.
\newblock In {\em Proceeding of the 2nd VLDB Conference on Secure Data
  Management}. Springer.
\newblock  185--199.

\bibitem[\protect\citeauthoryear{Bohn \bgroup et al\mbox.\egroup
  }{2005}]{bohn2005social}
Bohn, J.; Coroam{\u{a}}, V.; Langheinrich, M.; Mattern, F.; and Rohs, M.
\newblock 2005.
\newblock Social, economic, and ethical implications of ambient intelligence
  and ubiquitous computing.
\newblock In {\em Ambient Intelligence}. Springer.
\newblock  5--29.

\bibitem[\protect\citeauthoryear{Cheng \bgroup et al\mbox.\egroup
  }{2011}]{cheng2011exploring}
Cheng, Z.; Caverlee, J.; Lee, K.; and Sui, D.~Z.
\newblock 2011.
\newblock {Exploring Millions of Footprints in Location Sharing Services.}
\newblock {\em Proceedings of ICWSM'11} 2011:81--88.

\bibitem[\protect\citeauthoryear{Chow and Mokbel}{2011}]{chow2011trajectory}
Chow, C.-Y., and Mokbel, M.~F.
\newblock 2011.
\newblock Trajectory privacy in location-based services and data publication.
\newblock {\em ACM SIGKDD Explorations Newsletter} 13(1):19--29.

\bibitem[\protect\citeauthoryear{Colombo \bgroup et al\mbox.\egroup
  }{2012}]{colombo2012you}
Colombo, G.; Chorley, M.~J.; Williams, M.~J.; Allen, S.~M.; and Whitaker, R.~M.
\newblock 2012.
\newblock {You are where you eat: Foursquare checkins as indicators of human
  mobility and behaviour}.
\newblock In {\em Proceedings of PerCom Workshops 2012},  217--222.
\newblock IEEE.

\bibitem[\protect\citeauthoryear{Cramer, Rost, and
  Holmquist}{2011}]{cramer2011performing}
Cramer, H.; Rost, M.; and Holmquist, L.~E.
\newblock 2011.
\newblock Performing a check-in: emerging practices, norms and'conflicts' in
  location-sharing using foursquare.
\newblock In {\em Proceedings of MobileHCI'11},  57--66.
\newblock ACM.

\bibitem[\protect\citeauthoryear{de Montjoye \bgroup et al\mbox.\egroup
  }{2013}]{de2013unique}
de~Montjoye, Y.-A.; Hidalgo, C.~A.; Verleysen, M.; and Blondel, V.~D.
\newblock 2013.
\newblock Unique in the crowd: The privacy bounds of human mobility.
\newblock {\em Scientific Reports} 3.

\bibitem[\protect\citeauthoryear{Dwork}{2008}]{dwork2008differential}
Dwork, C.
\newblock 2008.
\newblock Differential privacy: A survey of results.
\newblock In {\em Theory and Applications of Models of Computation}. Springer.
\newblock  1--19.

\bibitem[\protect\citeauthoryear{Friedland and
  Sommer}{2010}]{friedland2010cybercasing}
Friedland, G., and Sommer, R.
\newblock 2010.
\newblock Cybercasing the joint: On the privacy implications of geo-tagging.
\newblock In {\em Proceedings of HotSec'10}.

\bibitem[\protect\citeauthoryear{Golle and
  Partridge}{2009}]{golle2009anonymity}
Golle, P., and Partridge, K.
\newblock 2009.
\newblock On the anonymity of home/work location pairs.
\newblock In {\em Pervasive Computing}. Springer.
\newblock  390--397.

\bibitem[\protect\citeauthoryear{Gruteser and
  Grunwald}{2003}]{gruteser2003anonymous}
Gruteser, M., and Grunwald, D.
\newblock 2003.
\newblock Anonymous usage of location-based services through spatial and
  temporal cloaking.
\newblock In {\em Proceedings of MobiSys'03},  31--42.
\newblock ACM.

\bibitem[\protect\citeauthoryear{Jin, Long, and Joshi}{2012}]{jin2012towards}
Jin, L.; Long, X.; and Joshi, J.~B.
\newblock 2012.
\newblock {Towards understanding residential privacy by analyzing users'
  activities in Foursquare}.
\newblock In {\em Proceedings of the 2012 ACM Workshop on Building Analysis
  Datasets and Gathering Experience Returns for Security},  25--32.
\newblock ACM.

\bibitem[\protect\citeauthoryear{Kalnis \bgroup et al\mbox.\egroup
  }{2007}]{kalnis2007preventing}
Kalnis, P.; Ghinita, G.; Mouratidis, K.; and Papadias, D.
\newblock 2007.
\newblock Preventing location-based identity inference in anonymous spatial
  queries.
\newblock {\em IEEE Transactions on Knowledge and Data Engineering}
  19(12):1719--1733.

\bibitem[\protect\citeauthoryear{Krumm}{2007}]{krumm2007inference}
Krumm, J.
\newblock 2007.
\newblock Inference attacks on location tracks.
\newblock In {\em Pervasive Computing}. Springer.
\newblock  127--143.

\bibitem[\protect\citeauthoryear{Krumm}{2009}]{krumm2009survey}
Krumm, J.
\newblock 2009.
\newblock A survey of computational location privacy.
\newblock {\em Personal and Ubiquitous Computing} 13(6):391--399.

\bibitem[\protect\citeauthoryear{Lindqvist \bgroup et al\mbox.\egroup
  }{2011}]{lindqvist2011m}
Lindqvist, J.; Cranshaw, J.; Wiese, J.; Hong, J.; and Zimmerman, J.
\newblock 2011.
\newblock I'm the mayor of my house: examining why people use foursquare-a
  social-driven location sharing application.
\newblock In {\em Proceedings of CHI'11},  2409--2418.
\newblock ACM.

\bibitem[\protect\citeauthoryear{Ma \bgroup et al\mbox.\egroup
  }{2013}]{ma2013privacy}
Ma, C. Y.~T.; Yau, D. K.~Y.; Yip, N.~K.; and Rao, N.~S.
\newblock 2013.
\newblock Privacy vulnerability of published anonymous mobility traces.
\newblock {\em IEEE/ACM Transactions on Networking} 21(3):720--733.

\bibitem[\protect\citeauthoryear{Manning, Raghavan, and
  Sch{\"u}tze}{2008}]{manning2008introduction}
Manning, C.~D.; Raghavan, P.; and Sch{\"u}tze, H.
\newblock 2008.
\newblock {\em Introduction to Information Retrieval}.
\newblock Cambridge University Press.

\bibitem[\protect\citeauthoryear{Narayanan and
  Shmatikov}{2008}]{narayanan2008robust}
Narayanan, A., and Shmatikov, V.
\newblock 2008.
\newblock Robust de-anonymization of large sparse datasets.
\newblock In {\em Proceedings of SP'08},  111--125.
\newblock IEEE.

\bibitem[\protect\citeauthoryear{Noulas \bgroup et al\mbox.\egroup
  }{2011}]{noulas2011empirical}
Noulas, A.; Scellato, S.; Mascolo, C.; and Pontil, M.
\newblock 2011.
\newblock An empirical study of geographic user activity patterns in
  foursquare.
\newblock {\em Proceedings of ICWSM'11} 11:70--573.

\bibitem[\protect\citeauthoryear{Noulas \bgroup et al\mbox.\egroup
  }{2012}]{noulas2012tale}
Noulas, A.; Scellato, S.; Lambiotte, R.; Pontil, M.; and Mascolo, C.
\newblock 2012.
\newblock A tale of many cities: universal patterns in human urban mobility.
\newblock {\em PloS one} 7(5):e37027.

\bibitem[\protect\citeauthoryear{Pontes \bgroup et al\mbox.\egroup
  }{2012a}]{pontes2012beware}
Pontes, T.; Magno, G.; Vasconcelos, M.; Gupta, A.; Almeida, J.; Kumaraguru, P.;
  and Almeida, V.
\newblock 2012a.
\newblock Beware of what you share: Inferring home location in social networks.
\newblock In {\em Proceedings of ICDM'12 Workshops},  571--578.
\newblock IEEE.

\bibitem[\protect\citeauthoryear{Pontes \bgroup et al\mbox.\egroup
  }{2012b}]{pontes2012we}
Pontes, T.; Vasconcelos, M.; Almeida, J.; Kumaraguru, P.; and Almeida, V.
\newblock 2012b.
\newblock {We Know Where you Live: Privacy Characterization of Foursquare
  Behavior}.
\newblock In {\em Proceedings of UbiComp'12},  898--905.
\newblock ACM.

\bibitem[\protect\citeauthoryear{Rossi and Musolesi}{2014}]{rossi2014s}
Rossi, L., and Musolesi, M.
\newblock 2014.
\newblock It's the way you check-in: identifying users in location-based social
  networks.
\newblock In {\em Proceedings of COSN'14},  215--226.
\newblock ACM.

\bibitem[\protect\citeauthoryear{Ruiz~Vicente \bgroup et al\mbox.\egroup
  }{2011}]{ruiz2011location}
Ruiz~Vicente, C.; Freni, D.; Bettini, C.; and Jensen, C.~S.
\newblock 2011.
\newblock Location-related privacy in geo-social networks.
\newblock {\em IEEE Internet Computing} 15(3):20--27.

\bibitem[\protect\citeauthoryear{Sweeney}{2002}]{sweeney2002k}
Sweeney, L.
\newblock 2002.
\newblock k-anonymity: A model for protecting privacy.
\newblock {\em International Journal of Uncertainty, Fuzziness and
  Knowledge-Based Systems} 10(05):557--570.

\bibitem[\protect\citeauthoryear{Zhao, Li, and Xue}{2013}]{zhao2013checking}
Zhao, X.; Li, L.; and Xue, G.
\newblock 2013.
\newblock Checking in without worries: Location privacy in location based
  social networks.
\newblock In {\em Proceedings of INFOCOM'13},  3003--3011.
\newblock IEEE.

\end{thebibliography}

\end{document}